\documentclass[pdftex,twocolumn,epjc3]{svjour3}

\usepackage[utf8]{inputenc}
\usepackage{amsmath}
\usepackage{amssymb}
\usepackage{graphicx}
\usepackage[colorlinks,citecolor=blue,urlcolor=blue,linkcolor=blue]{hyperref}
\usepackage[export]{adjustbox}
\usepackage[usenames,dvipsnames,svgnames,table]{xcolor}
\usepackage{multirow}

\newcommand{\de}{\delta}
\newcommand{\eps}{\varepsilon}
\newcommand{\eref}[1]{Eq. (\ref{#1})}
\newcommand{\fref}[1]{Fig. \ref{#1}}
\newcommand{\tref}[1]{Tab.~\ref{#1}}
\newcommand{\nnnl}{\nonumber\\}	

\allowdisplaybreaks 

\journalname{Eur. Phys. J. C}

\begin{document} 

\title{Higher spin glueballs from functional methods}

\author{Markus Q.~Huber\thanksref{e1,addr1} \and Christian S. Fischer\thanksref{e2,addr1,addr2} \and H\`elios Sanchis-Alepuz\thanksref{e3,addr3}}

\thankstext{e1}{e-mail: markus.huber@physik.jlug.de}
\thankstext{e2}{e-mail: christian.fischer@theo.physik.uni-giessen.de}
\thankstext{e3}{e-mail: helios.sanchis-alepuz@silicon-austria.com}

\institute{Institut f\"ur Theoretische Physik, Justus-Liebig-Universit\"at Giessen, Heinrich-Buff-Ring 16, 35392 Giessen, Germany\label{addr1}
          \and
          Helmholtz Forschungsakademie Hessen f\"ur FAIR (HFHF), GSI Helmholtzzentrum f\"ur Schwerionenforschung, Campus Gie{\ss}en, 35392 Gie{\ss}en, Germany\label{addr2}
          \and
          Silicon Austria Labs GmbH, Inffeldgasse 33, 8010 Graz, Austria\label{addr3}
}

\date{\today}

\maketitle

\begin{abstract}
We calculate the glueball spectrum for spin 
up to $J=$ 4 and positive charge parity in pure Yang-Mills theory.
We construct the full bases for $J=$ 0,1,2,3,4 and discuss the relation to gauge invariant operators.
Using a fully self-contained truncation of Dyson-Schwin\-ger equations as input, we obtain ground states and first and second excited states from extrapolations of the eigenvalue curves.
Where available, we find good quantitative agreement with lattice results.
\end{abstract}

\section{Introduction}

Hadrons that strictly consist of only gluons may not exist in nature. Instead, one expects mixtures of 
these so-called glueballs with corresponding meson states with the same quantum numbers. Nevertheless, 
it is important to study the spectrum of glueballs in pure Yang-Mills theory, since it may very well be
that some of these states obtain only small corrections from the matter sector of QCD. In this respect, it is 
very interesting to note that the scalar glueball predicted by lattice Yang-Mills theory long ago 
\cite{Bali:1993fb,Morningstar:1999rf,Chen:2005mg,Athenodorou:2020ani} and recently also within functional
methods \cite{Huber:2020ngt} seems to show up in radiative $J/\Psi$-decays \cite{Sarantsev:2021ein} with
almost unchanged mass. This merits further investigation in approaches that can deal with the (anti-)quark
admixtures of full QCD such as unquenched lattice calculations \cite{Gregory:2012hu}, the functional 
approach \cite{Huber:2020ngt,Dudal:2010cd,Dudal:2013wja,Meyers:2012ka,Sanchis-Alepuz:2015hma,Souza:2019ylx,Kaptari:2020qlt}, 
Hamiltonian many body methods \cite{Szczepaniak:1995cw,Szczepaniak:2003mr} or chiral Lagrangians
\cite{Janowski:2011gt,Eshraim:2012jv}, see also \cite{Klempt:2007cp,Crede:2008vw,Mathieu:2008me,Ochs:2013gi,Llanes-Estrada:2021evz} for reviews.  
     
In a previous work \cite{Huber:2020ngt}, we provided first results for the masses of ground and excited glueball states in the
scalar and pseudoscalar channels of pure Yang-Mills theory using a functional approach based on a fully
self-contained truncation of Dyson-Schwin\-ger equations and a set of Bethe-Salpeter equations derived from 
a three-particle-irreducible (3PI) effective action. In this work, we generalize our framework and present
additional results for ground and excited states with quantum numbers $J=2,3,4$ and positive as well as 
negative parity.

The obtained results are completely model independent in the sense that there are no free parameters which need to be tuned.
For the input, the only truncation appears at the level of the 3PI effective action which is truncated to three loops and allows a self-consistent determination of all the required correlation functions.
This feature of our calculation is different from other functional calculations which rely on phenomenological motivated model interactions.
Conceptually, it is appealing since extensions like the inclusion of quarks are naturally contained in this framework.
Technically, it comes at the price of more demanding calculations to obtain the input.

The manuscript is organized as follows. We first introduce the bound state equation in the next section, 
followed by a discussion of quantum numbers and how to construct bases for the Bethe-Salpeter amplitudes 
in Sec.~\ref{sec:bases}. The input and the procedure to obtain the glueball spectrum are explained in 
Sec.~\ref{sec:input} and the results are presented in Sec.~\ref{sec:results}.
We conclude with a summary in Sec.~\ref{sec:summary}.

\section{Glueball bound state equations}
\label{sec:bse}

\begin{figure*}[tb]
	\includegraphics[width=0.98\textwidth]{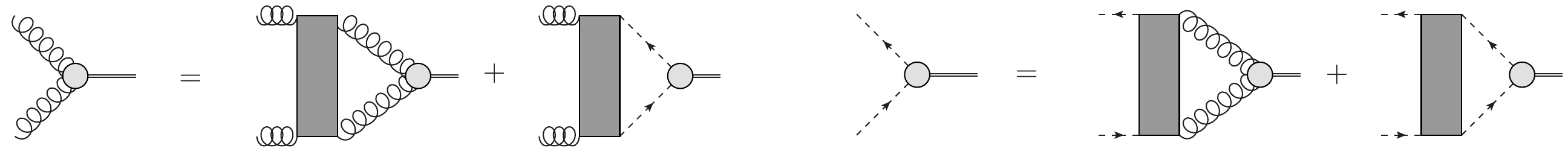}
	\caption{
		The coupled set of BSEs for a glueball made from two gluons and a pair of Faddeev-Popov (anti-)ghosts.
		Wiggly lines denote dressed gluon propagators, dashed lines denote dressed ghost propagators. 
		The gray boxes represent interaction kernels given in Fig.~\ref{fig:kernels}. The Bethe-Salpeter amplitudes of the glueball
		are denoted by gray disks. \label{fig:bses}
	}
	\vspace*{6mm}
	\includegraphics[width=0.48\textwidth]{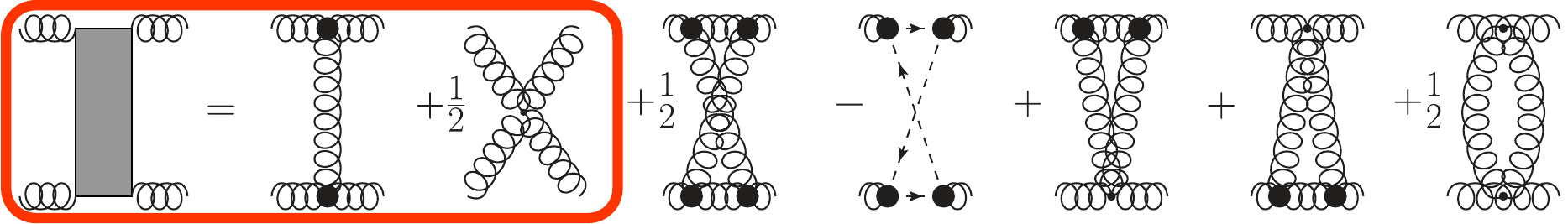}\\ 
	\vskip4mm
	\includegraphics[height=1.2cm]{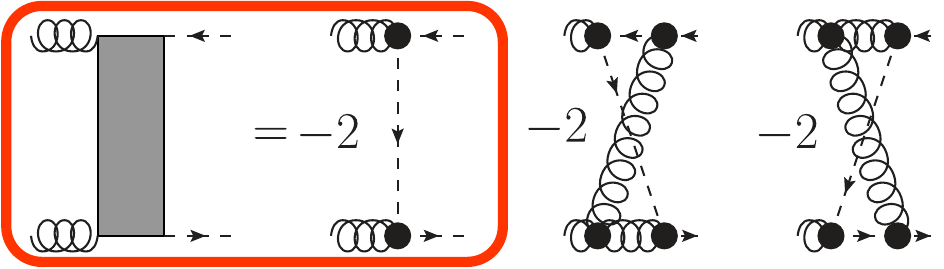}\hfill
	\includegraphics[height=1.2cm]{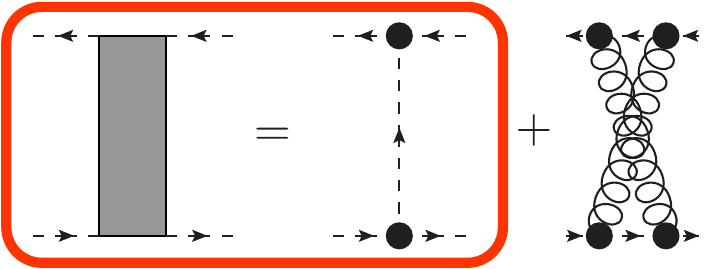}\hfill
	\includegraphics[height=1.2cm]{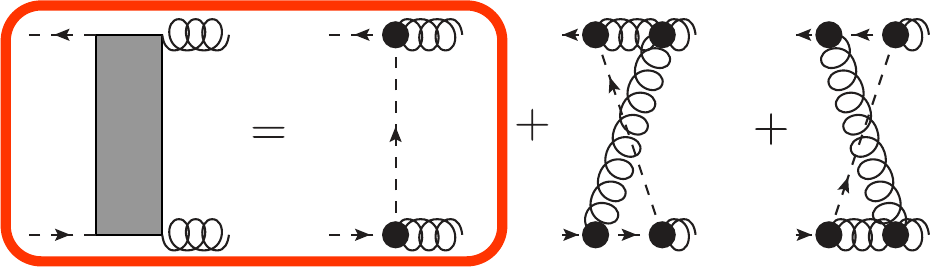}
	\caption{
		Interaction kernels from the three-loop 3PI effective action.
		All propagators are dressed; black disks represent dressed vertices.
		In our calculation, we include the diagrams inside the red rectangles. \label{fig:kernels}
	}
\end{figure*}

The properties of particles, elementary or composite, can be extracted from correlation functions.
We describe here how this is done for glueballs and how the Bethe-Salpeter formalism is formally related to the calculation of glueballs from gauge invariant operators on the lattice.
For pure (anti-)quark bound states, this is discussed, e.g., in \cite{Eichmann:2016yit}.
For glueballs, we have to pay particular attention to the fact that individual two- or three-gluon states themselves cannot be gauge invariant.
This is obvious when considering the operators used in lattice calculations to calculate glueball masses.
On the other hand, the functional formalism provides a means to extract the gauge invariant mass of bound states from gauge variant $n$-point functions.
This is particularly convenient for glueballs, where we can then get such information from the two-body bound state equations alone.

The properties of a particle can be extracted from an appropriate two-point function.
As a prime example for glueballs let us consider the composite operator $O_1(x)=F_{\mu\nu}(x) F^{\mu\nu}(x)$.
It has the quantum numbers $J^{\mathsf{PC}}=0^{++}$ \cite{Jaffe:1985qp}.
In momentum space, the two-point function $D(P)$,
\begin{align}
 &\langle 0| O_1(x) O_1(y) | 0 \rangle=D(x-y)=\nnnl
 &\quad \int \frac{d^4P}{(2\pi)^4}D(P)e^{-i\,(x-y)P},
\end{align}
has a pole at the mass of the $0^{++}$ glueball.
This is, for example, exploited in lattice calculations where the mass of a particle can be extracted from the exponential decay of such a two-point correlator at large times:
\begin{align}
 \lim_{t\rightarrow \infty}\langle O(x)O(0) \rangle \sim e^{-t\,M}.
\end{align}

Since the operator $O_1$ is gauge invariant, also the two-point function and the masses are gauge invariant.
Expanding the operator in gluon fields, one can write down a functional equation \cite{Pawlowski:2005xe}.
In the present example, this leads to an expression with up to six loops where all quantities are fully dressed.
The derivation is fully wor\-ked out for a similar example in Ref.~\cite{Haas:2014th} and can also be done 
algorithmically with a computer \cite{Huber:2019dkb}.
$O_1$ can be split into contributions with two, three, and four gluon fields, and one can group their contributions to $D(P)$ accordingly into terms with four, five, \ldots, eight gluon fields. A crucial point in determining the mass of a state is that its pole already shows up in one or more of the individual contributions, which therefore may be studied separately.

The one-loop contribution to $D(P)$ stems entirely from the two-gluon contribution to $O_1$, i.e. the correlation 
function of four gluons $G_{\mu\nu\rho\sigma}$:
\begin{align}
G_{\mu\nu\rho\sigma}(x,x;y,y)=\langle 0| A_\mu (x) A_\nu(x) A_\rho(y) A_\sigma(y)| 0\rangle.
\end{align}
Color indices are suppressed. This is a special case of the general four-gluon correlator $G_{\mu\nu\rho\sigma}(x,y,z,u)$,
from which we can derive a two-body bound state equation.
This Bethe-Salpeter equation (BSE) allows the determination of the pole mass corresponding to the scalar glueball.

This contrasts the calculation of glueball masses in the functional bound state and the lattice formalisms.
The latter uses correlation functions of gauge invariant operators like $O_1$, whereas the former relies on four-point functions of gluons which are gauge variant objects.
The information of the pole, however, is gauge invariant information contained in the four-point function.
The relation of composite operators like $O_1(x)$ to the tensor bases of Bethe-Salpeter amplitudes is explored in Sec.~\ref{sec:comp_ops}.

We close this discussion by emphasizing that the presence of a pole in the four-point function 
$G_{\mu\nu\rho\sigma}(x,y,z,u)$ is a sufficient condition for the presence of a pole in the correlator 
$D(P)$ assuming there are no cancellations with higher orders. While we cannot formally rule this out, 
in practice it is difficult to imagine how such a cancelation would come about even at the perturbative 
level. On the other hand, the absence of a pole in the four-point functions does not exclude the possibility 
of poles in $D(P)$, since it can still be present in higher contributions.

The Bethe-Salpeter equation for glueballs, shown in \fref{fig:bses}, couples a contribution containing two gluons (which we call glueball-part) and one containing a ghost and an antighost (ghostball-part). The ghost fields appear in the gauge fixing procedure; here we work in Landau gauge.
The corresponding kernels in the BSE have been derived from the 3PI effective action expanded to three 
loops \cite{Fukuda:1987su,McKay:1989rk,Berges:2004pu,Carrington:2010qq,Sanchis-Alepuz:2015tha} and
are shown in \fref{fig:kernels}. The derivation is detailed in Ref.~\cite{Huber:2020ngt}. The necessary 
input, i.e. the two and three-body correlators appearing in the kernels, is also obtained from this 
effective action \cite{Huber:2020keu}. In our main analysis, we only take into account the diagrams leading to one-loop expressions in the BSE. These are enclosed in the red boxes.
The inclusion of two-loop diagrams has to be deferred to future work, as they are computationally much more expensive.
For now, we take the quantitative agreement of our one-loop results with lattice results as indication that these diagrams are dominant.
Note also that decays are not possible in this setup as it would require additional diagrams, see, e.g., \cite{Fischer:2007ze}.

\section{Quantum numbers and tensor bases}
\label{sec:bases}

In this section, we explain which quantum numbers we calculate and derive the corresponding tensor bases.

\subsection{Landau-Yang theorem}\label{sec:LY}

Landau and Yang determined the allowed values for spin and parity quantum numbers of a system consisting of two photons \cite{Landau:1948kw,Yang:1950rg}.
They concluded that $J=1$ in general and $\mathsf{P}=-1$ for odd spin are forbidden.
This is often directly transferred to glueballs 'consisting of two gluons'.
We emphasize that one cannot construct a gauge invariant operator from two gluon fields and in this sense no pure two-gluon glueballs exist.
Here we consider the two-gluon part of the full gauge-invariant equation which can contain a pole by itself.
Below we discuss why the Landau-Yang theorem does not apply here and does not remove the pole from the two-gluon part.

The Landau-Yang theorem in the context of non-Abelian gauge theories was already considered in Refs.~\cite{Beenakker:2015mra,Cacciari:2015ela,Pleitez:2015cpa,Pleitez:2018lct}.
The decisive difference of QCD to QED is the color degree of freedom which leads to more allowed quantum numbers for color antisymmetric states.
However, such states are not relevant for the case of two gluons in a color singlet state.

Of direct relevance to our framework is one particular assumption that enters in the derivation of the Landau-Yang theorem, namely that the photons/gluons are on-shell.
The derivation proceeds by considering the parity even and odd cases separately.
In the former case, the state $J^\mathsf{P}=1^+$ is removed from the possible states by the on-shell condition.
For the parity odd case, the states $J^\mathsf{P}=(2k+1)^-$, with $k$ a nonnegative integer, are removed also by the on-shell condition and the fact that the spatial dimension is three \cite{Landau:1948kw,Pleitez:2015cpa}.
Since the gluons in the glueball are not on-shell, the on-shell condition is violated and the $J=1$ states are not removed.
Neither are the ones with odd $J$ and negative parity.
Thus, the Landau-Yang theorem does not forbid glueballs with these quantum numbers in our formalism.
Whether corresponding states actually appear as solutions of the two-body BSE is entirely a dynamical question.
We will see below, that in the current truncation scheme this is not the case. 

\subsection{Charge parity}

As described above, we employ a two-body bound state equation to calculate glueball masses.
This gives access to states with charge conjugation parity $\mathsf{C}=+1$ as we explain below.

Since the gluon field itself carries a color charge, it transforms nontrivially under charge conjugation.
The $SU(3)$ color matrices $\lambda^a$ get transposed under a charge conjugation.
Based on their symmetry properties, the gluon field transforms as \cite{Peccei:1998jv}
\begin{align}
 A^a_\mu\rightarrow -\eta(a)A^a_\mu
\end{align}
where
\begin{align}
 \eta(a)=\left\lbrace
        \begin{array}{l l}
          +1 & \quad a=1,3,4,6,8\\
          -1 & \quad a=2,5,7.
         \end{array}\right.
\end{align}
A color neutral operator with two gluon fields thus transforms as
\begin{align}
 A^a_\mu A^a_\nu \rightarrow \eta(a)^2 A^a_\mu A^a_\nu = A^a_\mu A^a_\nu.
\end{align}
This corresponds to positive charge parity.
Negative charge parity, on the other hand, can be realized, for example, with three gluon fields:
\begin{align}
 d^{abc} A^a_\mu A^b_\nu A^c_\rho \rightarrow& -d^{abc} \eta(a)\eta(b)\eta(c) A^a_\mu A^b_\nu A^c_\rho =\nnnl
 &- d^{abc}  A^a_\mu A^b_\nu A^c_\rho.
\end{align}
The last equality follows from the fact that the only nonvanishing elements of the symmetric structure constant $d^{abc}$ have zero or two indices equal to 2, 5 or 7.
For the antisymmetric structure constant, only the elements with one or three indices equal to 2, 5 or 7 are nonzero.
Consequently, this would lead to positive parity.

\subsection{Spin and parity}
\label{sec:spin_parity}

For the tensor bases we need to find all tensors compatible with a given spin $J$ and parity $\mathsf{P}$.
In addition, the transversality of the gluon propagator in Landau gauge needs to be taken into account.
We describe how to construct such tensor bases and work them out explicitly for $J=0,1,2,3,4$.

For glueballs, we only need to consider integer spin states.
Spin $J$ can then be described by a tensor $T$ of rank $J$.
It must be symmetric, traceless and transverse with respect to the total momentum $P$ \cite{Fierz:1939zz}:
\begin{subequations}
\label{eq:spinProjectorProperties}
\begin{align}
\label{eq:P_symmetry}
 T_{\ldots \mu\nu \ldots} & = T_{\ldots \nu\mu \ldots}, & \text{(symmetric)}\\
\label{eq:P_tracelessness}
 T_{\mu\mu\ldots}&=0, & \text{(traceless)}\\
\label{eq:P_transversality}
 P_{\mu}T_{\mu\ldots}&=0. & \text{(transverse)}
\end{align}
\end{subequations}
The second and third conditions remove lower spin states from the tensor and we end up with $2J+1$ independent elements \cite{Fierz:1939zz}.
These properties can be enforced on any tensor using appropriate spin projection operators $\mathcal{P}^J$, see, e.g., \cite{Behrends:1957rup}.
Let us consider as an example the case of spin two for which the projection operator reads
\begin{align}
\label{eq:proj_J2}
 \mathcal{P}^{J=2}_{\rho\sigma\rho'\sigma'}(P)&=\frac{1}{2}(\mathcal{P}_{\rho\rho'}(P)\mathcal{P}_{\sigma\sigma'}(P)+\mathcal{P}_{\rho\sigma'}(P)\mathcal{P}_{\sigma\rho'}(P))\nnnl
 &\quad -\frac{1}{3}\mathcal{P}_{\rho\sigma}(P)\mathcal{P}_{\rho'\sigma'}(P),
\end{align}
where
\begin{align}
 \mathcal{P}(k)_{\alpha\beta}=g_{\alpha\beta}-\frac{k_\alpha k_\beta}{k^2}
\end{align}
is the transverse projector.
Applying $\mathcal{P}^{J=2}$ on any tensor $\widetilde{T}_{\rho'\sigma'}$, the result obeys the conditions (\ref{eq:spinProjectorProperties}): It is symmetric in the two indices $\rho$ and $\sigma$, the trace is zero as ensured by the last term and it is transverse with respect to the total momentum $P$.

For the spin-2 example we can now easily construct a basis for the ghostball-part. We need tensors with two 
Lorentz indices. Using the metric tensor, the total momentum $P$ and the relative momentum $p$ we obtain 
five different structures. The transversality condition (\ref{eq:P_transversality}) leaves us with only 
$p_\mu p_\nu$ and $g_{\mu\nu}$. The tracelessness condition (\ref{eq:P_tracelessness}) further eliminates the latter.
Thus, the only admissible tensor for the spin-2 ghostball-part is
\begin{align}
 \lambda^\text{gh}_{\rho\sigma}=\mathcal{P}^{J=2}_{\rho\sigma\rho'\sigma'}p_{\rho'}p_{\sigma'}.
\end{align}
For other spins, the ghostball-part tensors are con\-struc\-ted in the same way by projecting $J$ relative momenta with a spin projector.

For the glueball-part, the procedure is similar, but now two additional Lorentz indices from the gluon legs enter the game.
For reference, we call them 'gluon leg indices' in contrast to the 'spin indices' $\rho$ and $\sigma$.
It is thus advantageous to filter the starting set of tensors appropriately, because for higher spin $J$, the number of possible tensors with $J+2$ indices increases quickly.
In the case of spin two, we can reduce the number of tensors constructed from the metric $g$ and two momenta from 43 to 10.
First of all, many tensors will not survive the spin projection due to the transversality or tracelessness conditions.
This can easily be taken into account by discarding all tensors with a spin index attached to a total momentum or metric tensors in the spin indices.
Furthermore, the symmetry condition entails that upon spin projection many tensors become linearly dependent.
To avoid that, we only take one representative for each case.
Following these selection rules, we determine for spin two the following initial set of tensors:
\begin{align}
\label{eq:init_basis_J2}
 \{& g_{\mu\rho}g_{\nu\sigma}, g_{\mu\nu} p_\rho p_\sigma, g_{\mu\rho} p_\nu p_\sigma, g_{\nu\rho} p_\mu p_\sigma, p_\mu p_\nu p_\rho p_\sigma;\nnnl
 &g_{\mu\rho} P_\nu p_\sigma, g_{\nu\rho} P_\mu p_\sigma, p_\mu P_\nu p_\rho p_\sigma, P_\mu p_\nu p_\rho p_\sigma, P_\mu P_\nu p_\rho p_\sigma \}.
\end{align}
$\mu$ and $\nu$ are the indices of the gluon legs and $\rho$ and $\sigma$ are spin indices. We call this set a 'pre-basis'.

\begin{figure}[tb]
 \begin{center}
 \includegraphics[width=0.38\textwidth]{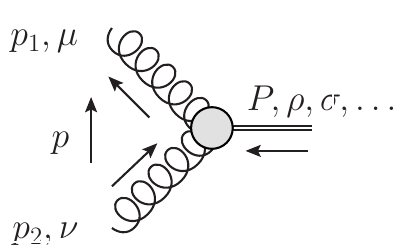}
 \end{center}
 \caption{Momenta and Lorentz indices of the glueball-part amplitude.}
 \label{fig:amplitude}
\end{figure}

As a next step, we consider the transversality of the gluon propagator in the Landau gauge.
It entails that the gluon legs of the glueball-part are transversely projected on the right-hand side of the BSE.
To be consistent with the left-hand side, the BSE amplitude should not contain nontransverse parts as otherwise the equation is no longer a proper eigenvalue equation.
Consequently, we need to modify the pre-basis such that it is invariant under $\mathcal{P}(p_1)_{\mu\mu'}\mathcal{P}(p_2)_{\nu\nu'}$,
where $p_1$ and $p_2$ are the momenta of the gluon legs, see \fref{fig:amplitude} for our naming conventions.
The gluon momenta are related to the total and relative momenta by
\begin{align}
\label{eq:pPp1p2}
 P=&p_1-p_2,\\
 p=&\frac{p_1+p_2}{2}.
\end{align}
Below we will use all four momenta $P,p,p_1,p_2$ for convenience, although of course only two of them are 
linearly independent.
A simple consequence of the transverse projection from the gluon propagators is that the momenta $P$ and $p$ 
attached to a gluon leg index are no longer independent and the pre-basis is reduced further.
In the example of spin two, one can choose the first five tensors in \eref{eq:init_basis_J2}.
These still need to be 'transversalized'. We illustrate this step explicitly in Sec.~\ref{sec:J0} 
for spin zero. Roughly speaking, it amounts to replacing momenta with a gluon leg index by a 
transversalized momentum. In our numerical treatment of the glueball BSE we find that the use of a 
transverse base is crucial to obtain correct results.

At this point we have what we call a 'transverse pre-basis'. To obtain the final bases, the spin projection 
operators $\mathcal{P}^{J}$ are applied to the elements of the pre-bases. This is discussed explicitly in
Secs.~\ref{sec:J0}--\ref{sec:J4}.

For negative parity states we additionally need the Levi-Civita symbol $\eps$ which requires a special discussion.
Due to its antisymmetric nature, it can appear only in a limited number of ways and we list the possible variants.
First, it can have either of the gluon indices $\mu$ or $\nu$.
Second, it can have only one index from the spin part because of the symmetric nature of the spin projector.
Third, it can have indices contracted with the relative or total momentum.
This leads to the following possibilities: 
\begin{align}
\eps_{\mu\nu\gamma\de}p_\gamma P_\de, \eps_{\mu\nu\rho\gamma}p_\gamma, \eps_{\mu\nu\rho\gamma}P_\gamma, \eps_{\mu\rho\gamma\delta}p_\gamma P_\delta, \eps_{\nu\rho\gamma\delta}p_\gamma P_\delta.
\end{align}
As stated above, any permutations of the spin indices are irrelevant because of the symmetry of the spin projector.

We now list the possible tensors for positive and negative parity.
In the former case, we take all tensors which can be constructed from the momenta and the metric $g$.
We do not need to consider an even number of Levi-Civita symbols $\eps$, as such an expression is linearly dependent on the previously mentioned terms.
This can be directly seen by writing the product of two Levi-Civita symbols in terms of metric tensors.

The following list of possible tensors for positive parity is exhaustive.
Further required indices are filled by appropriate expressions as discussed below.
The symbol '$\circ$' denotes a Lorentz index from the spin part.
\begin{itemize}
 \item Two metric tensors: There is only one tensor with $g_{\mu\circ}g_{\nu\circ}$.
 \item One metric tensor: There are three possibilities, name\-ly $g_{\mu\nu}$, $g_{\mu \circ}$ or $g_{\nu \circ}$.
 \item No metric tensor: There is exactly one tensor constructed from momenta only.
\end{itemize}
At this point we can already count the number of tensors for each $J$.
To this end, we observe that spin indices need to come with the relative momentum $p$ due to the transversality of the spin projector.
Open gluon leg indices, on the other hand, can come with two different momenta which, however, are linearly dependent when transversely projected.
Thus, every tensor will be represented only once.
We obtain two tensors for $J=0$ (of the types $g_{\mu\nu}$ and $p_\mu p_\nu$), four for $J=1$ (the type $g_{\mu\circ}g_{\nu\circ}$ does not exist) and five for higher spin (all possibilities realized).

For negative parity we have the following possibilities:
\begin{itemize}
 \item Both gluon leg indices in $\eps$: $\eps_{\mu\nu\alpha\beta}\,p_\alpha P_\beta$, $\eps_{\mu\nu\rho\alpha}\,p_\alpha$, $\eps_{\mu\nu\rho\alpha}\,P_\alpha$.
 \item One gluon leg index in $\eps$ and none in $g$: $\eps_{\mu\rho\alpha\beta}\,p_\alpha P_\beta p_\nu$, $\eps_{\nu\rho\alpha\beta}\,p_\alpha P_\beta p_\mu$.
 \item One gluon leg index in $\eps$ and one in $g$: $\eps_{\mu\rho\alpha\beta}\,p_\alpha P_\beta g_{\nu\sigma}$, $\eps_{\mu\rho\alpha\beta}\,p_\alpha P_\beta g_{\mu\sigma}$.
\end{itemize}
The tensors $\eps_{\mu\rho\alpha\beta}\,p_\alpha P_\beta P_\nu$, $\eps_{\nu\rho\alpha\beta}\,p_\alpha P_\beta P_\mu$ are not included, because they are linearly dependent after transverse projection of the gluon legs.
Counting the possible numbers of tensors taking into account the transversality of the gluon propagators leads to one for $J=0$ (of the type $\eps_{\mu\nu\alpha\beta}\,p_\alpha P_\beta$), five for $J=1$ (of the types $\eps_{\mu\nu\alpha\beta}\,p_\alpha P_\beta$, $\eps_{\mu\nu\rho\alpha}\,p_\alpha$, $\eps_{\mu\nu\rho\alpha}\,P_\alpha$, $\eps_{\mu\rho\alpha\beta}\,p_\alpha P_\beta p_\nu$,\linebreak $\eps_{\nu\rho\alpha\beta}\,p_\alpha P_\beta p_\mu$) and seven for higher spin.
However, these numbers reduce as some tensors are linearly dependent.
This can be shown by an explicit calculation.
The final numbers are then three for $J=1$ and four for higher spin.
We note that there is no ghostball-part for negative parity, because in that case we only have spin indices at our disposal of which we cannot put more than one into the Levi-Civita symbol.

We will now list the pre-bases for the glueball-parts for different spins constructed from the arguments above.
We also symmetrize the expressions with respect to exchanging the two legs.
For all basis tensors, the spin can be inferred from the number of indices.
Tensors with/without tilde refer to positive/negative parity.

\subsection{$J=0$ glueballs}
\label{sec:J0}

The scalar and pseudoscalar glueballs naturally have the simplest basis.
We are going to illustrate the step from the initial basis to the pre-basis, the transversalization, with the scalar one.
For positive parity, one can write down five tensors following the arguments outlined above:
\begin{align}
 g_{\mu\nu},&\quad& p_\mu p_\nu, & &\nnnl
 p_\mu P_\nu, &\quad& P_\mu p_\nu, &\quad& P_\mu P_\nu.
\end{align}
Taking into account the transversality of the gluon propagator, the last four tensors are equivalent, since
$$\mathcal{P}(p_1)_{\mu\mu'}P_{\mu'} \propto \mathcal{P}(p_1)_{\mu\mu'}p_{\mu'}$$
and similarly for the other leg with index $\nu$.
There are, thus, only two tensors in the transverse part.
Naively, one could take the tensors from the first line, but they are not invariant under a transverse projection of the gluon legs.
One could achieve this by simply projecting transversely.
Instead, we are going to use the auxiliary tensor
\begin{align}
 t^{ij}_{\mu\nu}&=g_{\mu\nu} p_i\cdot p_j -p_{j,\mu}p_{i,\nu},
\end{align}
where $i,j=1,2$.
For $i=1$, it is transverse in $p_{1,\mu}$ and similar for $j=2$ in $p_{2,\nu}$.
Two transverse tensors constructed from $t^{ij}_{\mu\nu}$ are then
\begin{subequations}
\label{eq:basisJ0}
\begin{align}
 \tau^1_{\mu\nu}&=\frac{t^{12}_{\mu\nu}}{p_1\cdot p_2}=g_{\mu\nu}-\frac{p_{2\mu}p_{1\nu}}{p_1\cdot p_2},\\
 \tau^2_{\mu\nu}&=\frac{t^{11}_{\mu\mu'}p_{2,\mu'}t^{22}_{\nu\nu'}p_{2,\nu'}}{(t^{11}p_2)\cdot(t^{22}p_1)}\nnnl
  &=\frac{(p_1^2 \,p_{2\mu}-p_1\cdot p_2\,p_{1\mu})(p_2^2\, p_{1\nu}-p_{1}\cdot p_2\, p_{2\nu})}{(p_1\cdot p_2)^3-p_1\cdot p_2\, p_1^2\, p_2^2}.
\end{align}
\end{subequations}
An appropriate normalization was chosen to make the tensors dimensionless.
This basis from Ref.~\cite{Meyers:2012ka} was employed in Ref.~\cite{Huber:2020ngt}.\footnote{We corrected here a typo for the momenta with index $\nu$.}
We also tried the variant based on $t^{11}_{\mu\alpha}t^{22}_{\alpha\nu}$, but it led to numerical instabilities.
The ghostball-part for the scalar glueball is simply a scalar.

For negative parity, only one tensor exists which can be chosen as
\begin{align}
 \widetilde{\tau}^1_{\mu\nu}&=\frac{\eps_{\mu\nu\rho\sigma}p_{\rho}P_{\sigma}}{\sqrt{p^2 P^2}}.
\end{align}
In Ref.~\cite{Huber:2020ngt} we used
\begin{align}
 \widetilde{\tau}^1_{\mu\nu}&=\frac{\eps_{\mu\nu\rho\sigma}\hat{p}^T_{\rho}\hat{P}_{\sigma}}{2}.
\end{align}
The hat indicates normalization and the superscript $T$ that the vector is made transverse with respect to $P$.
However, the two expressions are equivalent except for the normalization, as also the first tensor is transverse in the gluon legs.
This can be seen from the antisymmetry property of the Levi-Civita symbol due to which it can only hold two linearly independent momenta.
From the transverse projectors of the gluon propagators thus only the metric part survives.

\subsection{$J=1$ glueballs}
\label{sec:J1}

For spin one, we construct a basis from the set of all tensors with three Lorentz indices.
As we saw for the scalar glueball, it is sufficient to consider only the relative momentum $p$ for the gluon leg indices due to the linear dependence introduced by the transverse projection.
For positive parity, we obtain then the tensors $g_{\mu\nu}p_{\rho}$, $g_{\mu\rho}p_{\nu}$, $g_{\nu\rho}p_{\mu}$, $p_{\mu}p_{\nu}p_{\rho}$.
We recombine them such that they are symmetric under exchange of the gluons and invariant under transverse projection of the gluon legs.
Furthermore, we normalize them conveniently.
First, however, it is useful to introduce the following auxiliary momenta:
\begin{subequations}
\begin{align}
 p_{1,\mu}^T=t^{11}_{\mu\mu'}p_{2,\mu'}=t^{11}_{\mu\mu'}p_{\mu'},\\
 p_{2,\nu}^T=t^{22}_{\nu\nu'}p_{1,\nu'}=t^{22}_{\nu\nu'}p_{\nu'}.
\end{align}
\end{subequations}
Our choice for the transverse pre-basis is
\begin{subequations}
\begin{align}
\tau^1_{\mu\nu\rho}&=\frac{t^{12}_{\mu\nu}\,p_{\rho}\,p\cdot P}{p_1\cdot p_2\, p^2\sqrt{P^2}},\\
\tau^2_{\mu\nu\rho}&=\frac{(t^{11}_{\mu\rho}\,p^T_{2,\nu}+t^{22}_{\nu\rho}\,p^T_{1,\mu})p\cdot P}{p_1^2\, p_2^2\, p^2\sqrt{P^2}},\\
\tau^3_{\mu\nu\rho}&=\frac{t^{11}_{\mu\rho}\,p^T_{2,\nu}-t^{22}_{\nu\rho}\,p^T_{1,\mu}}{p_1^2\, p_2^2\, \sqrt{p^2}},\\
\tau^4_{\mu\nu\rho}&=\frac{p^T_{1,\mu}p^T_{2,\nu}p_{\rho}\,p\cdot P}{p_1^2 p_2^2 p^4\sqrt{P^2}}.
\end{align}
\end{subequations}
To single out the contributions relevant for spin one, we apply the spin-1 projector, which is identical to the transverse projector with respect to the total momentum:
\begin{align}
 \mathcal{P}^{J=1}_{\rho\rho'}=\mathcal{P}_{\rho\rho'}(P)=g_{\rho\rho'}-\frac{P_\rho P_\rho'}{P^2}.
\end{align}

For negative parity, the following tensors can be constructed:
\begin{subequations}
\begin{align}
 \widetilde{\tau}^1_{\mu\nu\rho}&=\frac{\eps_{\mu\nu\gamma\delta}\,p_{\gamma}\,P_\delta\, p_\rho\,p\cdot P}{(p^2)^\frac{3}{2}P^2},\\
 \widetilde{\tau}^2_{\mu\nu\rho}&=\frac{t^{11}_{\mu\alpha}t^{22}_{\nu\beta}\eps_{\alpha\beta\rho\gamma}\,p_{\gamma}}{p_1^2\,p_2^2\,\sqrt{p^2}},\\
 \widetilde{\tau}^3_{\mu\nu\rho}&=\frac{t^{11}_{\mu\alpha}t^{22}_{\nu\beta}\eps_{\alpha\beta\rho\gamma}\,P_{\gamma}\,p\cdot P}{p_1^2\,p_2^2\,\sqrt{p^2}P^2},\\
 \widetilde{\tau}^4_{\mu\nu\rho}&=\frac{\left(\frac{\eps_{\mu\rho\gamma\delta}\,p^T_{2,\nu}}{p_2^2}+\frac{\eps_{\nu\rho\gamma\delta}\,p^T_{1,\mu}}{p_1^2}\right)\,p_{\gamma}\,P_\delta}{p^2\sqrt{P^2}},\\
 \widetilde{\tau}^5_{\mu\nu\rho}&=\frac{\left(\frac{\eps_{\mu\rho\gamma\delta}\,p^T_{2,\nu}}{p_2^2}-\frac{\eps_{\nu\rho\gamma\delta}\,p^T_{1,\mu}}{p_1^2}\right)\,p_{\gamma}\,P_\delta\, p\cdot P}{p^\frac{3}{2}P^2}.
\end{align}
\end{subequations}
After spin-1 projection, only three of these tensors are linearly independent.
The first three tensors are a possible choice for the pre-basis.

\subsection{$J=2$ glueballs}
\label{sec:J2}

The initial tensors for spin two were discussed in Sec.~\ref{sec:spin_parity}.
Following the transversalization procedure outlined above, we obtain the following transverse pre-basis:
\begin{subequations}
\label{eq:basis_J2}
\begin{align}
\tau^1_{\mu\nu\rho\sigma}&=\frac{t^{12}_{\mu\rho}t^{21}_{\nu\sigma}}{(p_1\cdot p_2)^2},\\
\tau^2_{\mu\nu\rho\sigma}&=\frac{t^{12}_{\mu\nu}\,p_{\rho}p_{\sigma}}{p_1\cdot p_2\,p^2},\\
\tau^3_{\mu\nu\rho\sigma}&=\frac{t^{12}_{\mu\rho}\,p^T_{2,\nu}\,p_{\sigma}}{p_1 \cdot p_2\, a_2\,\sqrt{p^2}}+\frac{t^{21}_{\nu\rho}\,p^T_{1,\mu}\,p_{\sigma}}{p_1 \cdot p_2\, a_1\,\sqrt{p^2}},\\
\tau^4_{\mu\nu\rho\sigma}&=\left(\frac{t^{12}_{\mu\rho}\,p^T_{2,\nu}\,p_{\sigma}}{p_1 \cdot p_2\, a_2}-\frac{t^{21}_{\mu\rho}\,p^T_{1,\nu}\,p_{\sigma}}{p_1 \cdot p_2\, a_1}\right)\frac{p\cdot P}{p^2\sqrt{P^2}},\\
\tau^5_{\mu\nu\rho\sigma}&=\frac{p^T_{1,\mu}\,p^T_{2,\nu}\,p_\rho\,p_\sigma}{a_1\,a_2\,p^2},
\end{align}
\end{subequations}
where
\begin{align}
a_i&=\sqrt{(p^T_i)^2}.
\end{align}
An appropriate normalization of all tensors was chosen for convenience.
Applying the spin-2 projector (\ref{eq:proj_J2}) yields the basis.

The possible tensors for negative parity are
\begin{subequations}
\begin{align}
\widetilde \tau^1_{\mu\nu\rho\sigma}&=\frac{\eps_{\mu\nu\gamma\delta}\,p_\gamma\,P_\delta\,p_\rho\,p_\sigma}{(p^2)^\frac{3}{2}\sqrt{P^2}},\\
\widetilde \tau^2_{\mu\nu\rho\sigma}&=\frac{t^{11}_{\mu\alpha}t^{22}_{\nu\beta}\eps_{\alpha\beta\rho\gamma}\,p_\gamma\,p_\sigma\,p\cdot P}{p_1^2 \,p_2^2\,(p^2)^\frac{3}{2}\sqrt{P^2}},\\
\widetilde \tau^3_{\mu\nu\rho\sigma}&=\frac{t^{11}_{\mu\alpha}t^{22}_{\nu\beta}\eps_{\alpha\beta\rho\gamma}\,P_\gamma\,p_\sigma}{p_1^2 \,p_2^2\,\sqrt{p^2\,P^2}},\\
\widetilde \tau^4_{\mu\nu\rho\sigma}&=\frac{\left(t^{11}_{\mu\alpha}\eps_{\alpha\rho\gamma\delta}\,p^T_{2,\nu}+t^{22}_{\nu\beta}\eps_{\beta\rho\gamma\delta}\,p^T_{1,\mu}\right)p_\gamma\,P_\delta\,p_\sigma\,p\cdot P}{p_1^2\,p_2^2\,p^4\, P^2},\\
\widetilde \tau^5_{\mu\nu\rho\sigma}&=\frac{\left(t^{11}_{\mu\alpha}\eps_{\alpha\rho\gamma\delta}\,p^T_{2,\nu}-t^{22}_{\nu\beta}\eps_{\beta\rho\gamma\delta}\,p^T_{1,\mu}\right)p_\gamma\,P_\delta\,p_\sigma}{p_1^2\,p_2^2\,(p^2)^\frac{3}{2}\,\sqrt{P^2}},\\
\widetilde \tau^6_{\mu\nu\rho\sigma}&=\frac{\left(t^{11}_{\mu\alpha}\eps_{\alpha\rho\gamma\delta}\,t^{22}_{\nu\sigma}+t^{22}_{\nu\beta}\eps_{\beta\rho\gamma\delta}\,t^{11}_{\mu\sigma}\right)p_\gamma\,P_\delta\,p\cdot P}{p_1^2\,p_2^2\,p^2\, P^2},\\
\widetilde \tau^7_{\mu\nu\rho\sigma}&=\frac{\left(t^{11}_{\mu\alpha}\eps_{\alpha\rho\gamma\delta}\,t^{22}_{\nu\sigma}-t^{22}_{\nu\beta}\eps_{\beta\rho\gamma\delta}\,t^{11}_{\mu\sigma}\right)p_\gamma\,P_\delta}{p_1^2\,p_2^2\,\sqrt{p^2\,P^2}}.
\end{align}
\end{subequations}
Of these seven tensors, four are linearly independent after transverse projection.
We choose the tensors $\widetilde{\tau}^1$, $\widetilde{\tau}^2$, $\widetilde{\tau}^3$, and $\widetilde{\tau}^6$.

\subsection{$J=3$ glueballs}
\label{sec:J3}

The bases for spin three can directly be obtained from those of $J=2$ by multiplying with additional vectors $p_\sigma$, correcting the power of $p\cdot P$ to enforce symmetry in the gluon legs and updating the normalization.
No new terms arise and since the procedure is straightforward we do not give the explicit expressions.

The projection operator for spin three is \cite{Huang:2005js}
\begin{align}
 \mathcal{P}&^{J=3}_{\rho\sigma\tau\rho'\sigma'\tau'}(P)=\frac{1}{6}\Big(\mathcal{P}_{\rho\rho'}(P)\mathcal{P}_{\sigma\sigma}(P)\mathcal{P}_{\tau\tau'}(P)\nnnl
 &+\mathcal{P}_{\rho\rho'}(P)\mathcal{P}_{\sigma\tau'}(P)\mathcal{P}_{\tau\sigma'}(P)+\mathcal{P}_{\rho\sigma'}(P)\mathcal{P}_{\sigma\rho'}(P)\mathcal{P}_{\tau\tau'}(P)\nnnl
 &+\mathcal{P}_{\rho\sigma'}(P)\mathcal{P}_{\sigma\tau'}(P)\mathcal{P}_{\tau\rho'}(P)+\mathcal{P}_{\rho\tau'}(P)\mathcal{P}_{\sigma\sigma'}(P)\mathcal{P}_{\tau\rho'}(P)\nnnl
 &+\mathcal{P}_{\rho\tau'}(P)\mathcal{P}_{\sigma\rho'}(P)\mathcal{P}_{\tau\sigma'}(P)\Big)\nnnl
 &-\frac{1}{15}\Big(\mathcal{P}_{\rho\sigma}(P)\mathcal{P}_{\tau\rho'}(P)\mathcal{P}_{\sigma'\tau'}(P)\nnnl
 &+\mathcal{P}_{\rho\sigma}(P)\mathcal{P}_{\tau\sigma'}(P)\mathcal{P}_{\rho'\tau'}(P)\nnnl
 &+\mathcal{P}_{\rho\sigma}(P)\mathcal{P}_{\tau\tau'}(P)\mathcal{P}_{\sigma'\tau'}(P)\nnnl
 &+\mathcal{P}_{\rho\tau}(P)\mathcal{P}_{\nu\rho'}(P)\mathcal{P}_{\sigma'\tau'}(P)\nnnl
 &+\mathcal{P}_{\rho\tau}(P)\mathcal{P}_{\nu\sigma'}(P)\mathcal{P}_{\rho'\tau'}(P)\nnnl
 &+\mathcal{P}_{\rho\tau}(P)\mathcal{P}_{\nu\tau'}(P)\mathcal{P}_{\sigma'\rho'}(P)\nnnl
 &+\mathcal{P}_{\nu\tau}(P)\mathcal{P}_{\rho\rho'}(P)\mathcal{P}_{\sigma'\tau'}(P)\nnnl
 &+\mathcal{P}_{\nu\tau}(P)\mathcal{P}_{\rho\sigma'}(P)\mathcal{P}_{\rho'\tau'}(P)\nnnl
 &+\mathcal{P}_{\nu\tau}(P)\mathcal{P}_{\rho\tau'}(P)\mathcal{P}_{\sigma'\rho'}(P)\Big).
\end{align}
An alternative notation, which is easier to generalize to spin four, reads
\begin{align}
 \mathcal{P}&^{J=3}_{\rho\sigma\tau\rho'\sigma'\tau'}=\frac{1}{(3!)^2}\sum_{\substack{\text{perm. } \rho,\sigma,\tau\\ \text{perm. }\rho',\sigma',\tau'}}\Big(\mathcal{P}_{\rho\rho'}(P)\mathcal{P}_{\sigma\sigma'}(P)\mathcal{P}_{\tau\tau'}(P)\nnnl
 &\quad -\frac{3}{5}\mathcal{P}_{\rho\sigma}(P)\mathcal{P}_{\rho'\sigma'}(P)\mathcal{P}_{\tau\tau'}(P)\Big).
\end{align}

\subsection{$J=4$ glueballs}
\label{sec:J4}

Again, only a new momentum $p_\zeta$ needs to be added to the $J=3$ tensors with resulting modifications for the power of $p\cdot P$ and the normalization.
The spin projector for $J=4$ is \cite{Huang:2005js}
\begin{align}
 \mathcal{P}&^{J=4}_{\rho\sigma\tau\zeta\rho'\sigma'\tau'\zeta'}(P)=\nnnl
 &\quad\frac{1}{(4!)^2}\sum_{\substack{\text{perm. } \rho,\sigma,\tau,\zeta\\ \text{perm. }\rho',\sigma',\tau',\zeta'}}\Big(\mathcal{P}_{\rho\rho'}(P)\mathcal{P}_{\sigma\sigma'}(P)\mathcal{P}_{\tau\tau'}(P)\mathcal{P}_{\zeta\zeta'}(P)\nnnl
 &\quad -\frac{6}{7}\mathcal{P}_{\rho\sigma}(P)\mathcal{P}_{\rho'\sigma'}(P)\mathcal{P}_{\tau\tau'}(P)\mathcal{P}_{\zeta\zeta'}(P))\nnnl
&\quad +\frac{3}{35}\mathcal{P}_{\rho\sigma}(P)\mathcal{P}_{\rho'\sigma'}(P)\mathcal{P}_{\tau\zeta}(P)\mathcal{P}_{\tau'\zeta'}(P)\Big).
\end{align}

\subsection{Composite operators for glueballs}
\label{sec:comp_ops}

\begin{figure*}[t]
	\includegraphics[width=0.48\textwidth]{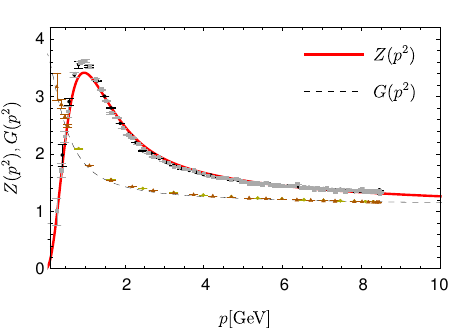}\hfill
	\includegraphics[width=0.48\textwidth]{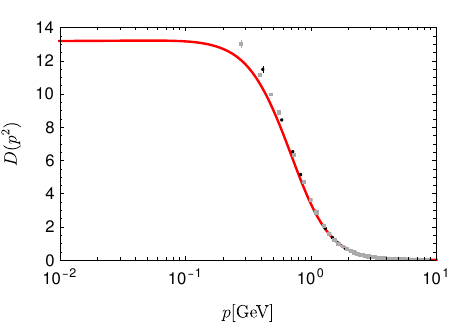}
	\caption{Gluon and ghost dressing functions $Z(p^2)$ and $G(p^2)$, respectively, (left) and gluon propagator $D(p^2)$ (right) in comparison to lattice data \cite{Sternbeck:2006rd}.
		For the sake of comparison, the functional results were renormalized (in the plots only) to agree with the lattice results at $8$ (ghost) and $6\,\text{GeV}$ (gluon).
	}
	\label{fig:props}
	\includegraphics[width=0.48\textwidth]{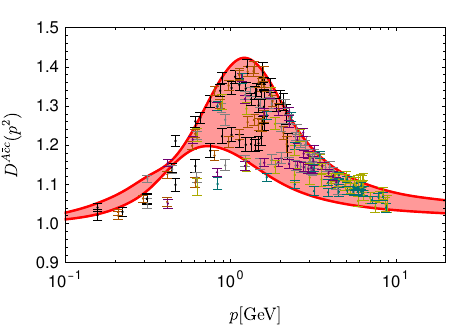}\hfill
	\includegraphics[width=0.48\textwidth]{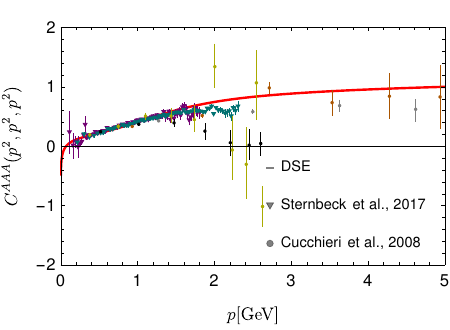}
	\caption{
		Left: Ghost-gluon vertex dressing function (full kinematic dependence) in comparison to $SU(2)$ lattice data \cite{Maas:2019ggf}.
		Right: Three-gluon vertex dressing function at the symmetric point in comparison to lattice data \cite{Cucchieri:2008qm,Sternbeck:2017ntv}), see Refs.~\cite{Athenodorou:2016oyh,Boucaud:2017obn} for similar results.
		For the sake of comparison, the three-gluon vertex data was renormalized (in the plot only) to $1$ at $5\,\text{GeV}$.
	}
	\label{fig:verts}
\end{figure*}

The bases we constructed can be related to the physical picture of glueballs described in form of gauge invariant operators as given, e.g., in Ref.~\cite{Jaffe:1985qp}.
Let us consider the following examples:
\begin{align}
    S&=\text{tr} F_{\mu\nu}F^{\mu\nu},& J^{\mathsf{PC}}=0^{++},\\
    P&=\text{tr} F_{\mu\nu}\widetilde{F}^{\mu\nu}, &J^{\mathsf{PC}}=0^{-+},\\
    T_{\rho\sigma}&=\text{tr} F_{\rho\alpha}F^\alpha_\sigma-\frac{1}{4}g_{\rho\sigma}S, &J^{\mathsf{PC}}=0^{++},1^{-+},2^{++}.
\end{align}
Evidently, these operators contain two-, three and four-gluon components. As discussed in Sec.~\ref{sec:bse}, poles in the correlation functions of any of these components correspond to glueball masses.
Since we consider a two-gluon bound state equation, we expect that the bases for the Bethe-Salpeter amplitudes have a correspondence in the two-gluon component of such gauge invariant operators.
This can be checked by applying two derivatives with respect to gluon fields and setting the fields to zero.
We exemplify this with the operator $T$, from which we obtain 10 terms:
\begin{subequations}
\label{eq:T_tensors}
\begin{align}
 g_{\mu\nu}g_{\rho\sigma} && && \rightarrow & \tau^1_{\mu\nu}g_{\rho\sigma}\quad  (J=0)\nnnl
 g_{\mu\rho}g_{\nu\sigma}, && 
 g_{\mu\sigma}g_{\nu\rho} && \rightarrow & \tau^1_{\mu\nu\rho\sigma}\quad  (J=2)\nnnl
 g_{\mu\nu}p_{1\rho}p_{2\sigma}, &&
 g_{\mu\nu}p_{2\rho}p_{1\sigma}  && \rightarrow & \tau^2_{\mu\nu\rho\sigma}\quad  (J=2)\nnnl
 g_{\mu\rho}p_{1\nu}p_{2\sigma}, &&
 g_{\mu\sigma}p_{1\nu}p_{2\rho}  && \rightarrow & \tau^{3,4}_{\mu\nu\rho\sigma}\quad  (J=2)\nnnl
 g_{\nu\rho}p_{2\mu}p_{1\sigma}, && 
 g_{\nu\sigma}p_{2\mu}p_{1\rho}  && \rightarrow & \tau^{3,4}_{\mu\nu\rho\sigma}\quad  (J=2)\nnnl
 g_{\rho\sigma}p_{2\mu}p_{1\nu} && && \rightarrow & \tau^2_{\mu\nu}g_{\rho\sigma}\quad  (J=0).
\end{align}
\end{subequations}
The first and last, which vanish upon spin-2 projection, are identified with a basis for the scalar glueball multiplied by $g_{\rho\sigma}$, see Sec.~\ref{sec:J0}.
Making these tensors transverse, relates them to the basis in \eref{eq:basisJ0}.
Some of the other tensors are identical upon projection with the spin-2 projector.
For instance, this applies to the second and third tensors.
Upon closer inspection, we find that tensors two and three correspond (after transversalization) to $\tau^1_{\mu\nu\rho\sigma}$, tensors four and five to $\tau^2_{\mu\nu\rho\sigma}$, and tensors six to nine to $\tau^3_{\mu\nu\rho\sigma}$ and $\tau^4_{\mu\nu\rho\sigma}$ of \eref{eq:basis_J2}.
The third column in \eqref{eq:T_tensors} indicates these correspondences.
The basis element $\tau^5_{\mu\nu\rho\sigma}$ does not appear here.
It is interesting to note that we find the amplitude of this tensor to be subleading.

\section{Solving the BSEs: input and extrapolation}
\label{sec:input}

\begin{table*}[tb]
	\begin{center}
		\begin{tabular}{|l||c|c|c|c|c|c|c|c|}
			\hline
			&  \multicolumn{2}{c|}{\cite{Morningstar:1999rf}} & \multicolumn{2}{c|}{\cite{Chen:2005mg}} & \multicolumn{2}{c|}{\cite{Athenodorou:2020ani}} & \multicolumn{2}{c|}{This work}\\   
			\hline
			State &  $M\, [\text{MeV}]$& $M/M_{0^{++}}$ & $M\, [\text{MeV}]$& $M/M_{0^{++}}$  &  $M\, [\text{MeV}]$& $M/M_{0^{++}}$ & $M\,[\text{MeV}]$ & $M/M_{0^{++}}$\\   
			\hline\hline
			$0^{++}$ & $1760 (50)$ & $1(0.04)$ & $1740(60)$ & $1(0.05)$ & $1651(23)$ & $1(0.02)$ & $1850 (130)$ & $1(0.1)$\\
			\hline
			$0^{^*++}$ & $2720 (180)^*$ & $1.54(0.11)^*$ & -- & -- & $2840(40)$ & $1.72(0.034)$ & $2570 (210)$ & $1.39(0.15)$\\
			\hline
			\multirow{2}{*}{$0^{^{**}++}$} & \multirow{2}{*}{--} & \multirow{2}{*}{--} & \multirow{2}{*}{--} & \multirow{2}{*}{--} & $3650(60)^\dagger$ & $2.21(0.05)^\dagger$ & \multirow{2}{*}{$3720 (160)$} & \multirow{2}{*}{$2.01(0.16)$}\\
			& & & & & $3580(150)^\dagger$ & $2.17(0.1)^\dagger$ & &\\
			\hline
			$0^{-+}$ & $2640 (40) $ & $1.50(0.05)$ & $2610(50)$ & $1.50(0.06)$ & $2600(40)$ & $1.574(0.032)$ & $2580 (180)$ & $1.39(0.14)$\\
			\hline
			$0^{^*-+}$ & $3710 (60)$ & $2.10(0.07)$ & -- & -- & $3540(80)$ & $2.14(0.06)$ & $3870 (120)$ & $2.09(0.16)$\\
			\hline
			\multirow{2}{*}{$0^{^{**}-+}$} & \multirow{2}{*}{--} & \multirow{2}{*}{--} & \multirow{2}{*}{--} & \multirow{2}{*}{--} & $4450(140)^\dagger$ & $2.7(0.09)^\dagger$ & \multirow{2}{*}{$4340 (200)$} & \multirow{2}{*}{$2.34(0.19)$}\\
			& & & & & $4540(120)^\dagger$ & $2.75(0.08)^\dagger$ & &\\
			\hline
			\hline
			$2^{++}$ & 2447(25) & 1.39(0.04) & 2440(50) & 1.40(0.06) & 2376(32) & 1.439(0.028) & 2610(180) & 1.41(0.14)\\
			\hline
			$2^{^*++}$ & -- & -- & -- & -- & 3300(50) & 2(0.04) & 3640(240) & 1.96(0.19)\\
			\hline
			$2^{-+}$ & 3160(31) & 1.79(0.05) & 3100(60) & 1.78(0.07) & 3070(60) & 1.86(0.04) & 2740(140) & 1.48(0.13)\\
			\hline
			$2^{^*-+}$ &  3970(40)$^*$ & 2.25(0.07)$^*$ & -- & -- & 3970(70) & 2.4(0.05) & 4300(190) & 2.32(0.19)\\
			\hline\hline
			$3^{++}$ & 3760(40) & 2.13(0.07) & 3740(60) & 2.15(0.09) & 3740(70)$^*$ & 2.27(0.05)$^*$ & 3370(50)$^*$ & 1.82(0.13)$^*$\\
			\hline
			$3^{^*++}$ & -- & -- & -- & -- & -- & -- & 3510(170)$^*$ & 1.89(0.16)$^*$\\
			\hline
			$3^{^{**}++}$ & -- & -- & -- & -- & -- && 3970(220)$^*$ & 2.14(0.19)$^*$\\
			\hline
			$3^{-+}$ & -- & -- & -- & -- & -- & -- & 4050(290)$^*$ & 2.19(0.22)$^*$\\
			\hline
			\hline
			$4^{++}$ & -- & -- & -- & -- & 3690(80)$^*$ & 2.24(0.06)$^*$ & 4140(30)$^*$ & 2.23(0.15)$^*$\\
			\hline
			$4^{-+}$ & -- & -- & -- & -- & -- & -- & 3240(300)$^*$ & 1.75(0.2)$^*$\\
			\hline
		\end{tabular}
		\caption{Ground and excited state masses $M$ of glueballs for various quantum numbers.
		Compared are lattice results from \cite{Morningstar:1999rf,Chen:2005mg,Athenodorou:2020ani} with the results of this work and \cite{Huber:2020ngt}.
        For \cite{Morningstar:1999rf,Chen:2005mg}, the errors are the combined errors from statistics and the use of anisotropic lattices.
		For \cite{Athenodorou:2020ani}, the error is statistical only.
		In our results, the error comes from the extrapolation method and should be considered a lower bound on errors.
		All results use the same value for $r_0=1/(418(5)\,\text{MeV})$.
		The related error is not included in the table.
		Masses with $^\dagger$ are conjectured to be the second excited states.
		Masses with $^*$ come with some uncertainty in their identification in the lattice case or in the trustworthiness of the extrapolated value in the BSE case.
		}
		\label{tab:masses}
	\end{center}
\end{table*}

We follow the same setup as in \cite{Huber:2020ngt} which we shortly summarize here.
To solve the BSE, we need the ghost and gluon propagators and the ghost-gluon and three-gluon vertices.
We take them from a parameter-free calculation \cite{Huber:2020keu}.
The corresponding quantities are compared to lattice results in Figs.~\ref{fig:props} and \ref{fig:verts}.
For a meaningful comparison, the functional results were renormalized to the values indicated in the captions.
For the calculations, the original data was used which was renormalized as explained in Ref.~ \cite{Huber:2020keu}.
They also agree very well with results from the functional renormalization group \cite{Cyrol:2016tym}.
We emphasize that a family of solutions can be obtained which all fulfill the standard Landau gauge fixing condition \cite{Boucaud:2008ji,Fischer:2008uz,Alkofer:2008jy,Maas:2009se,Maas:2011se,Sternbeck:2012mf,Huber:2018ned,Eichmann:2021zuv}.
These solutions may correspond to different nonperturbative gauge completions of the perturbative Landau gauge \cite{Maas:2009se}.
In Ref.~\cite{Huber:2020ngt} we tested the dependence of our results for $J=0$ on the different solutions.
We found that for all solutions, including the one with a diverging ghost dressing function, that the obtained masses agree within the errors induced by the extrapolation.
Hence, we continue here with only one solution, the one shown in Figs.~\ref{fig:props} and \ref{fig:verts}.

It is worth mentioning that these results are independent of the number of colors as the system scales trivially with $g^2\,N_c$.
For Yang-Mills theory, nontrivial terms only enter at four loops \cite{vanRitbergen:1997va}.
The truncation in \cite{Huber:2020ngt} does not capture these terms.
As the employed BSE also scales trivially in $g^2\,N_c$, our results are hence independent of $N_c$.

Since the two- and three-point functions necessary as input for the BSEs are only available for Euclidean momenta, 
we cannot solve the BSE directly at the time-like pole-locations, $-M^2=P^2<0$. To access time-like momenta, we 
instead extrapolate the eigenvalue curve from the space-like side using  Schles\-sin\-ger's method based on 
continued fractions \cite{Schlessinger:1968spm,Tripolt:2018xeo}. We tested the reliability of the eigenvalue 
extrapolation by calculating a meson with a setup that allows calculations for time-like momenta \cite{Huber:2020ngt}.
In that case, the method was very accurate for masses up to roughly 2~GeV. Beyond that, the errors increase.
For the present case, where no direct estimation of the extrapolation error is possible, we estimate it indirectly 
by taking 100 samples of 80 points out of 100 calculated points for the extrapolation. From this we calculate the 
error indicated in the results section. 
The error estimated from this method establishes only a lower bound and results for large masses should be 
interpreted with care.

The identification of states is also somewhat involved for several reasons. First, it is not necessary that 
the hierarchy of eigenvalues at space-like momenta agrees with the hierarchy in the physical region, since eigenvalue 
curves may cross at time-like $P^2$. It is then not clear, whether the extrapolation method is still reliable.
Second, we also encountered cases where it was not possible to identify a unique eigenvalue curve with a state, 
because there are several solutions leading to similar masses. We indicate all such cases in the discussion in the 
next section.

For higher spin states the numerical effort to determine excited states increases drastically due to the 
larger number of Lorentz indices. As a consequence we had to reduce the numerical accuracy in order to comply with
the available CPU time. We checked the effect of this on the scalar and pseudoscalar glueballs and found that 
the ground states are unaffected but excited states may disappear with decreased accuracy. This explains why 
we do not always find two excited states for $J>0$, as discussed below.
For spin three and four and positive parity, we also did not take into account the diagrams containing ghosts to reduce the computational cost.
However, we confirmed for one eigenvalue that the amplitude from the ghostball-part is subleading compared to the ones from the glueball-part.

\section{Results}
\label{sec:results}

\begin{figure*}[tb]
	\begin{center}
	\includegraphics[width=0.68\textwidth]{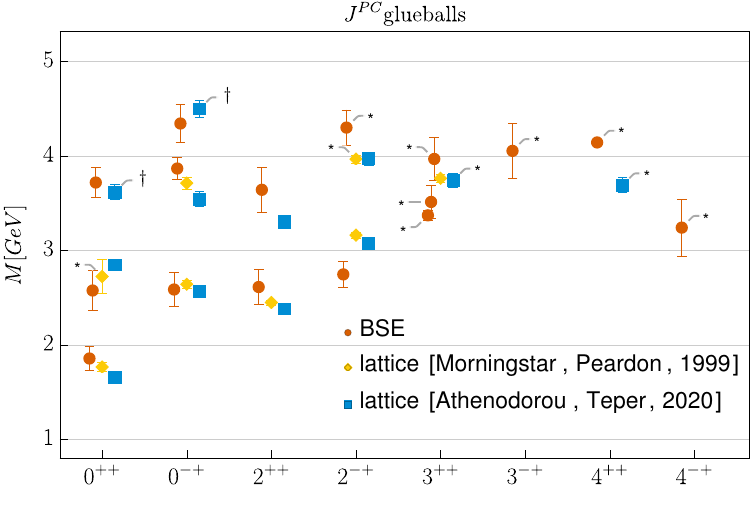}\\
	\includegraphics[width=0.68\textwidth]{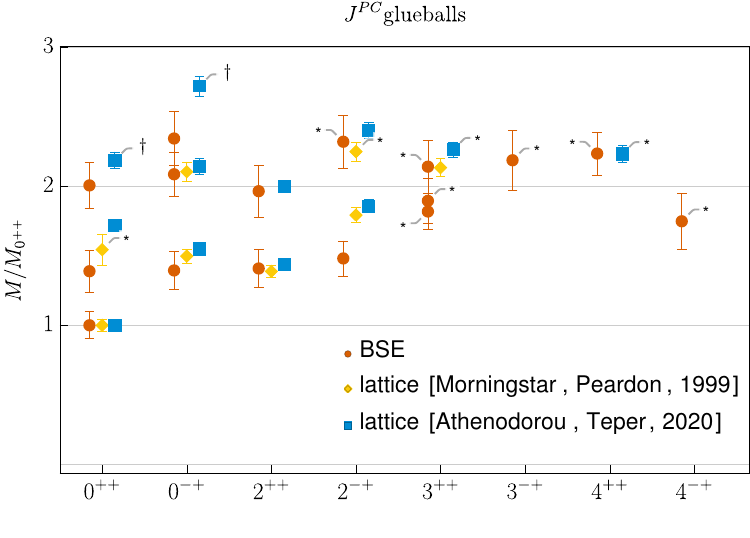}
	\end{center}
	\caption{
		Results for glueball ground states and excited states for the indicated quantum numbers from lattice simulations \cite{Morningstar:1999rf,Athenodorou:2020ani} and functional equations.
		In the upper plot, we display the glueball masses on an absolute scale set by $r_0=1/(418(5)\,\text{MeV})$.
	    In the lower plot, we display the spectrum relative to the ground state.
	    Masses with $^\dagger$ are conjectured to be the second excited states.
		Masses with $^*$ come with some uncertainty in their identification in the lattice case or in the trustworthiness of the extrapolated value in the BSE case.
		}
	\label{fig:spectrum}
\end{figure*}

Our results for glueball states with $J=0,\ldots,4$ and positive and negative parity are summarized in \fref{fig:spectrum} 
and \tref{tab:masses}.
For the comparison with lattice results, we rescale all results to the same value $r_0=0.472(5)\,\text{fm}$ of the Sommer scale used in \cite{Athenodorou:2020ani}, see Ref.~\cite{Huber:2020ngt} for details.

The results for $J=0$ are taken from Ref.~\cite{Huber:2020ngt} and are unambiguous.
We found clear signals for the ground states and two excited states which compare well with lattice results.
Note that the second excited states on the lattice were not uniquely identified in Ref.~\cite{Athenodorou:2020ani}.
In fact, for $\mathsf{P}=+1$ and for $\mathsf{P}=-1$ two states were found in each case with very similar masses, see \tref{tab:masses}.
The good agreement with our results indicate that one of them is most likely indeed a second excited $J=0$ state.

The identification of states for $J=2$ is more challenging than for spin zero. For $2^{++}$, we observed a degeneracy 
in the ground state and first excited masses, viz., we found two eigenvalue curves for each that yield, within 
extrapolation errors, the same masses. The curves themselves, however, are not degenerate. Inspecting the Bethe-Sal\-peter
amplitudes from each of these two solutions, we identified a crucial difference: one set does not show the 
power law fall-off in the large momentum region as expected for a normalizable Bethe-Salpeter amplitude. We adopted this 
as a criterion to rule out solutions of this type, as discussed already in Ref.~\cite{Huber:2020ngt}. Within error bars,
the two remaining solutions for the ground and first excited states are in very good agreement with corresponding lattice results.
We also found a candidate for a second excited state at 3.8\,GeV, but the error from the extrapolation is more than 1\,GeV in that case so we refrain from including it in the table.

Our results for the masses in the $2^{-+}$ channel again show an interesting pattern with seemingly two 
solution branches. Inspecting the amplitudes of one branch we find a clear pattern with respect to zero crossings: 
no crossings for the ground state and $n$ zero-crossings for the $n$-th excited state. The amplitudes furthermore 
resemble qualitatively those of the pseudoscalar glueball. This pattern is not seen, however, for the second branch 
of solutions, which we therefore discard as artifact of the truncation. In the physical branch, we find an excited 
state in the mass region beyond 4\,GeV. Since its extraction involves quite some extrapolation into the time-like region, its mass value comes with a large uncertainty.

For glueballs of spins three and four, we obtain masses between 3 and 4.5\,GeV.
Again, we assign some additional uncertainty to these values due to the extrapolation.
Let us first discuss the positive parity states.
For $3^{++}$, we find two excited states. All three states are rather close to each other. 
It remains to be seen whether this still persists in an improved approach where we would be able to follow the eigenvalue curves into the time-like region. In principle, we cannot exclude that two of the curves merge at some point and disappear
(i.e. the eigenvalues become complex). Such a behaviour has already been seen in approaches with simpler truncations, 
where time-like momenta are accessible, see e.g. \cite{Eichmann:2016yit,Eichmann:2016hgl} for an overview. In our case, this would mean
that not all three states would survive. In contrast, our ground state for $4^{++}$ can be identified exceptionally cleanly
and is in the same range as the corresponding lattice result.
The extrapolation is extremely stable in that case what leads to the unusually small error.
The reason for that is currently unknown but we also observed it for some other eigenvalues.
As far as lattice results are concerned, we also have to keep in mind that the identification of the $3^{++}$ and $4^{++}$ states is not as unique as for the lighter states \cite{Athenodorou:2020ani}. 
The negative parity states $3^{-+}$ and $4^{-+}$ have not been identified on the lattice and we regard our results
as predictions, albeit with large uncertainties, especially since the spin four state is lighter than the spin three state.

Finally, we also solved the BSE for spin one. As discussed in section \ref{sec:LY}, there is no principal reason why such a state should not appear in a two-body bound state equation.
However, dimensional counting of the corresponding operators containing
these quantum numbers suggests that the masses should be large \cite{Morningstar:1999rf,Kuti:1998rh}. Indeed, 
in the mass region considered in this work we were not able to find solutions fulfilling the criteria for the amplitudes discussed above. For $1^{++}$, this agrees with the results from lattice Yang-Mills theory, while for $1^{-+}$ a mass of about 4\,GeV was found \cite{Athenodorou:2020ani}.

\section{Summary and discussion}
\label{sec:summary}

We calculated the spectrum of glueballs with $\mathsf{C}=1$ and $J=0,1,2,3,4$ from a two-body bound state equation.
For $J=1$, we did not find solutions. For the other cases, we were able to find the ground states and up to 
two excited states.
The results for $J=0$ have been discussed already previously in Ref.~\cite{Huber:2020ngt}.
A crucial factor to arrive at these results was the use of high-quality input for the gluon and ghost 
correlation functions from a self-contained calculation \cite{Huber:2020keu}. As the only parameter of the 
input is the scale, this is a calculation from first principles.
Our results agree qualitatively and quantitatively with corresponding lattice results
\cite{Morningstar:1999rf,Chen:2005mg,Athenodorou:2020ani}, especially when comparing the ratios to the lightest state.
This is very encouraging.
The largest deviations we observe for the $2^{-+}$ and $3^{++}$ groundstates.

The present setup can be improved in several ways. An obvious one concerns the inclusion of the neglected 
two-loop diagrams in the BSE to establish its self-con\-sis\-tency. This is computationally very costly but 
in principle possible. However, due to the good agreement with lattice results so far, we do not expect 
large corrections at least for the lighter states. More importantly, we rely on an extrapolation of the 
eigenvalue curves from space-like to time-like momenta. A calculation directly at $P^2=-M^2$ could shed 
light on some of the uncertainties we encountered in the identification of physical states and remove 
the extrapolation error. Unfortunately, for such a calculation we require two- and three-point functions 
evaluated in the time-like (and complex) momentum regime, which is currently not available in sufficient 
quality. Some corresponding calculations exist but are less advanced in their truncations
\cite{Strauss:2012dg,Fischer:2020xnb,Horak:2021pfr}.

Of course, the most obvious and interesting extension of the present framework concerns the inclusion of
the matter sector of QCD. This is work in progress.

\section*{Acknowledgments}

This work was supported by the DFG (German Research Foundation) grant FI 970/11-1 and by the BMBF under 
contracts No. 05P18RGFP1 and 05P21RGFP3.
This work has also been supported by Silicon Austria Labs (SAL), owned by the Republic of Austria, the Styrian Business Promotion Agency (SFG), the federal state of Carinthia, the Upper Austrian Research (UAR), and the Austrian Asso­ci­a­tion for the Elec­tric and Elec­tronics Industry (FEEI).

\bibliographystyle{utphys_mod}
\bibliography{literature_glueballs_J2}

\end{document}